\documentclass{ijuc}
\usepackage{graphicx}

\begin{document}

\title{An Overview: Steady-State Quantum
	Entanglement via Reservoir
	Engineering}

\author{Ali Pedram\inst{1}\email{apedram19@ku.edu.tr}
	\and \"{O}zg\"ur E. M\"{u}stecapl{\i}o\u{g}lu\inst{1,2}\email{omustecap@ku.edu.tr}
}

\institute{Department of Physics, Ko\c{c} University, Sar{\i}yer, Istanbul, 34450, T\"{u}rkiye
\and
T\"{U}B\.{I}TAK Research Institute for Fundamental Sciences, 41470 Gebze, T\"{u}rkiye
}

\maketitle

\begin{abstract}
We present a short overview of quantum entanglement generation and preservation in a steady state. In addition to the focus on quantum entanglement stabilization, we briefly discuss the same objective for steady-state quantum coherence. The overview classifies the approaches into two main categories: hybrid drive and dissipation methods and purely dissipative schemes. Furthermore, purely dissipative schemes are discussed under two subclasses of equilibrium and nonequilibrium environments. The significance of the dissipative route to
sustained quantum entanglement and challenges against it are pointed out. Besides the value of steady-state entanglement for existing quantum technologies, quantum computation, communication, sensing, and simulation, its unique opportunities for emerging and future quantum technology applications, particularly quantum heat engines and quantum energy processing, are discussed.
\end{abstract}


\section{Introduction}

One of the weird aspects of quantum entanglement which puzzles and fascinates people is that it can be envisioned as a link between objects independent of their distance. The non-local, instantaneous, and distance-free influence of entangled particles upon each other is confused even the greatest minds of physics, such as Albert Einstein, who coined the term “spooky action at a distance” to the quantum entanglement. Nowadays, modern quantum technologies benefit a lot from the practical advantages of the non-local ubiquitous character of entangled particles, which gives unique opportunities for information teleportation channels exploited in the field of quantum communication. Surprisingly, the potential benefits of quantum
non-locality in space are limited by the fragile nature of quantum entanglement in time. Setting quantum entanglement on initially distant particles is not trivial. Typically, one would entangle particles nearby then distribute them, which requires protection of quantum entanglement during the transportation of the particles. Even if the particles are kept locally, quantum entanglement tends to dissipate into surrounding environments. Long-time sustainability of quantum entanglement, either locally or during transport, especially when many entangled particles are exposed to the environment, is one of the critical challenges of practical quantum technologies, such as long-time ultra-precise quantum measurements, quantum memories for quantum computation, and long-distance quantum communication.

An experimentally available strategy to sustain quantum entanglement in the steady-state is to exploit decoherence-free subspaces (DFS's)~\cite{Kwiat2000}, in which the environmental dissipation can be used to confine the dynamics~\cite{Beige2000}. Alternative methods include dynamical decoupling or Zeno effect schemes, where coherent pulse sequences are applied to reverse the information loss ~\cite{Viola1998,DeLange2010}; and measurement-feedback procedures, for continuous monitoring and correcting the quantum state to be protected~\cite{Wiseman1993,Riste2013,Riste2015}. Search for quantum entanglement protection with less control, complexity, and resources led to the hybrid coherent drive and engineered dissipation strategies and, more recently, purely dissipative approaches.

In this short overview, we present the key contributions and developments, to our knowledge, to generate and protect quantum entanglement in the steady-state. In addition, we briefly mention similar efforts for sustaining quantum coherence using dissipation. Our overview categorizes the contributions to the field into two broad classes: the schemes include coherent drives and the methods without coherent drives. The latter has two sub-categories: procedures with multiple baths or a single bath. The common understanding of the advantages of dissipative schemes to produce and protect quantum coherence and entanglement is motivated mainly by quantum error correction applications, which benefit from initial state independent steady-state and its relatively more straightforward scalability. Our discussion has another, perhaps a subjective, energy perspective beyond these usual benefits of dissipative methods. Our energy perspective includes reducing the energetic footprint of quantum technologies and opening unprecedented quantum technology fields, such as quantum heat management (diodes, transistors, switches), quantum heat, and information engines.

\section{Hybrid Coherent Drive and Engineered Dissipation Schemes To Sustain Quantum Entanglement}

A common strategy to preserve quantum entanglement over time is not much different than sustaining oscillations in a damped resonator by applying a resonant driving sinusoidal force. Aron, Kulkarni, and T\"ureci, for example, proposed to generate photon-mediated interactions among many ($N$) two-level atoms (qubits) in the resonator(s) using external coherent drives to counterbalance their quantum decoherence through resonator damping, qubit dephasing and dissipative emission~\cite{Aron2016}.
Careful selection of the coherent drive parameters and locations of the atoms in the shared electromagnetic field environment allows for engineering effective qubit-qubit coupling models, an analog of magnetic, interacting spin models, with
the desired quantum entangled steady-states occupied with the help of the dissipative dynamics. The method can be used for dissipative generation of scaled $N$-qubit entangled states.

Kimchi-Schwartz and co-workers present an experimental implementation of a coherent drive and bath engineering strategy by using superconducting flux-tunable transmon qubits coupled to copper waveguide cavities~\cite{Kimchi-Schwartz2016}. By utilizing spatial symmetry of the bath as a novel degree of freedom in bath engineering, the authors show that using only a single microwave drive is sufficient to counterbalance the quantum decoherence to produce and maintain two-qubit entangled Bell states in steady-state.

Steady-state quantum entanglement is naturally advantageous as it is present even in experimental noise and environmental quantum decoherence. In addition to this arbitrary-time existence benefit, steady-state quantum entanglement brings another opportunity to reduce quantum error correction overheads since it is independent of the initial state and hence is less error-prone relative to completely unitary coherent schemes. Nevertheless,
hybrid strategies combine unitary evolution with engineered dissipative dynamics, and thus they still have
coherent components in their implementations, which
generally require careful and precise control of multiple external coherent drives. Coherent drives, especially in continuous, autonomous approaches, consume significant energy
to protect the entanglement over a long time as the external drives must always be on to counterbalance the quantum decoherence. Autonomous quantum entanglement generators and protectors require minimal or no external control. For example, Shankar and co-workers experimentally demonstrated that two superconducting qubits could be evolved into a Bell state, preserved for an arbitrary amount of time, by continuous coherent drives and engineered dissipation in an autonomous feedback scheme~\cite{Shankar2013}. A similar approach has resulted in the production and preservation of two-qubit Bell states with trapped ions, too~\cite{Lin2013}. Leghtas and co-workers specifically discussed and proposed an autonomous quantum memory for quantum computing architectures requiring less error-correcting qubits~\cite{Leghtas2013}.

Preservation of large entangled systems remains a critical experimental challenge. Intriguingly, the scaling of entanglement can be envisioned in two aspects. First, the entangled objects can have a large size or mass, yet the number of entangled massive or sizeable objects can still be small. The second aspect of scaling is to have tiny quantum objects, like electronic spins or quantum particles, but the number of entangled particles could be considerable. The latter is the more relevant objective for practical quantum technologies. The former is significant for ultrasensitive quantum measurements, such as gravity sensing, fundamental questions of the quantum collapse of wavefunctions, tests of quantum limits, and explorations of the boundary between classical and quantum realms. There are promising contemporary experiments that achieved sustainable entanglement of a pair of massive objects. In 2018, Aolto university researchers demonstrated that two massive and macroscopic (barely visible to the naked eye) aluminum disks (drum heads) could be entangled in a steady-state using microwave cavity photons to synchronize their vibrations~\cite{Ockeloen-Korppi:2018tg}. The mechanical entanglement between massive oscillators persists for an arbitrary time as long as the microwave cavity is coherently and strongly pumped. More recently, measurement methods have been improved for continuously monitoring macroscopic entanglement without destroying it~\cite{Mercier-de-Lepinay:2021vx}. It seems increasing the number of entangled objects, even if they are tiny qubits, is more challenging than increasing the mass or size of two entangled objects in terms of decoherence.

Quantum decoherence may scale with the number of qubits as well. One can think that an increase in the number of entangled qubits could bring a superlinear quantum advantage (for large
number ($N$) qubits, there are about $N^2$ possible entanglement connections among them). On the other hand, if the decoherence increase similarly with the system size (such as with $N^2$), then the qubits would have entanglement in steady-state up to a critical number of qubits or system size, after which the decoherence would remove any quantum superiority. In terms of energetics of quantum entanglement, such a quantum law of diminishing returns has been
established in Ref.~\cite{Hardal:2018wa}. Quantum decoherence channels may be less forgiving in scaling with mass and size in the entanglement of two macroscopic mechanical oscillators~\cite{Ockeloen-Korppi:2018tg}, though one still relies on powerful coherent drives to fight against them. Another pioneering experiment related to the steady-state entanglement of macroscopic objects via drive and dissipation methods uses large atomic ensembles~\cite{Krauter:2011tl}, where the entanglement is obtained in continuous variables and is of EPR type, particularly useful for quantum metrology and sensing.

Remarkably, the experimental success to achieve steady-state quantum entangled states even with a pair of qubits are not as perfect as the theoretical predictions. The difference is due to unaccounted quantum dephasing and experimental noise effects reducing the experimental fidelity. Hence, scaling to a more significant number of qubits in experiments and making the experimental progress more promising for practical quantum error correction that needs stabilization of quantum entanglement with many qubits, the question of increasing fidelity for multi-qubit entanglement is critical.
Hein, Aron, and T\"ureci have proposed that stable, and higher fidelity quantum entanglement can still be obtained without introducing much additional complexity and coherent resources via hybrid drive-dissipation schemes~\cite{Hein2016}.
Hence, we have experimental evidence and solid theoretical arguments including scalability to multiple qubits that the precise control or many coherent drive requirements can be mitigated by using clever bath engineering~\cite{Kimchi-Schwartz2016} and autonomous operation schemes to get high fidelity steady-state quantum entanglement in the presence of experimental noise, decoherence, and dephasing.

A remaining challenge is how to reduce the energy footprint of such quantum entanglement generators and protectors.
Energy perspective and energetic constraints are usually not considered as of high priority when designing quantum technologies, though recent studies remark the significance of these questions for scalable and practical quantum devices in the near future~\cite{Auffeves2021,Fellous-Asiani2021}.

\section{Completely Incoherent Schemes To Sustain Quantum Enganlement}

Using only thermal, incoherent resources to generate and sustain quantum entanglement has recently attracted much attention. In these methods, the system to be entangled is treated as an open quantum system. The methods can be classified into two broad categories. First, the open quantum system can be in nonequilibrium environments; for example, two-coupled qubits attached to two environments at different temperatures or chemical potentials. The energy flow between the reservoirs through the qubit system can power the quantum entanglement. These studies, however, usually neglect other, uncontrolled, local environments of the
qubits that may be present in practice. The second class contains schemes that rely on only a single environment, for example, two coupled qubits immersed in a shared thermal bath. The latter is called a thermal entangled system. Thermal entanglement can be potentially valuable for practical quantum devices that can operate at higher temperatures~\cite{Hsiang:2015vd};. However, other decoherence channels, such as dephasing, could destroy thermal entanglement quickly, and thermal environment per se may not be sufficient to sustain entanglement in practice~\cite{Znidaric:2012tj}. Z.~Wang and co-workers present a detailed study of steady-state quantum coherence and entanglement of two-coupled qubits in equilibrium and nonequilibrium environments~\cite{Wang:2019th}. An earlier study explores similar scenarios, emphasizing quantum thermalization in independent and shared baths, but also concludes that there are threshold temperatures for qubits to be entangled in thermal environments~\cite{Liao:2011uc}.

\subsection{Sustained Quantum Entanglement in Nonequilibrium Environments}

When interacting qubits open to environments, quantum entanglement that can emerge out of the interaction can be expected to vanish. On the other hand, if there is a constant energy flow through the qubit system between the
surrounding non-equilibrium environments, the entanglement can survive~\cite{Quiroga:2007tx}. Early studies are limited to thermal environments for entanglement of few qubits ~\cite{Sinaysky:2008wv,Hu:2018wc,Valle:2011vl,Duan:2013wb,Scala:2008vv} or quantum coherence generation in few level~\cite{Li:2015ua} or few qubit~\cite{Huangfu:2017tq} systems.
For example, X.~L.~Huang and co-workers studied thermal entanglement among
three qubits between two thermal baths at different temperatures~\cite{Huang:2009wn}. Contrary to the possible expectation, the optimum conditions for higher entanglement are not always when the heat flow is maximum or when the temperatures are highly
different. The symmetric couplings between the qubits require an equilibrium bath configuration where the baths are at the same temperatures. If the qubit-qubit couplings are not the same, such a nonsymmetric interacting system is entangled more when the baths are at more different temperatures. In addition to the symmetries and interaction structures, the strength of interaction between the qubits, relative to the energy flow, plays a decisive role in the fate of thermally sustained quantum entanglement. The effects of so-called entanglement sudden death~\cite{Yu:2004wk,Ann:2007wu} and birth can emerge at those non-equilibrium quantum open systems. A particular scenario is a case where two qubits are coupled to a larger qubit chain carrying energy flow, studied in Ref.~\cite{Liu:2009up}. Depending on the coupling strength between the qubits and the energy flow, entanglement can abruptly disappear but then revive after some time, albeit at a smaller amount.

Other examples of sustained entanglement in open quantum systems in nonequilibrium environments can also be found in electronic transport problems. The different tunneling rates of electric charges to their electronic environments ensure the nonequilibrium transport conditions thorough two Coulomb-coupled double quantum dots for the entanglement of the electric charges~\cite{Lambert:2007vs}.
Hybrid systems have been considered for steady-state entanglement with nonequilibrium environments, too.
L.~D.~Contreras-Pulido and co-workers theoretically investigated steady-state entanglement of two semiconducting charge-qubits, whose interaction is mediated by the photons of a microwave cavity hosting the double quantum dots containing the charge-qubits~\cite{Contreras-Pulido:2013vn}.

Sustaining quantum coherence for arbitrary times has been experimentally demonstrated using a superconducting qubit subject to continuous drive and dissipation in the dispersive regime of circuit QED, where the intensity (photon number) of the superconducting transmission line resonator drives the qubit polarization.

An Environment can either be local to an individual qubit or
shared by several or all qubits in a system. Depending on the local or shared nature of the baths,
they can play different, positive, or negative roles in sustaining the quantum entanglement. While local baths can cause an abrupt loss of quantum entanglement (entanglement sudden death), shared baths can revive the entanglement (entanglement sudden birth) (For a review, see Ref.~\cite{Hernandez:2010wc}). Z.-X. Man and co-workers showed that a shared bath for two-qubits, next to their local baths, yield more quantum entanglement in the steady-state~\cite{Man:2019tc}.

As a digression here, let us briefly address the generation and protection of quantum coherence. One may think that even multiple qubits or spins can easily be polarized or aligned in one direction using external drives or high magnetic fields, and hence the coherence generation and sustainability are trivial. However, one should be careful that quantum coherence is basis dependent quantum resource. One can, for example, simply align a spin qubit along an axis determined with the orientation of an applied magnetic field and sustain the spin alignment for an arbitrary time as long as the magnetic field is present. Such a spin would be in a quantum superposition (coherent) state with respect to any perpendicular axes to the alignment axis. In the second quantum revolution, it is desired to develop devices or applications that harness quantum superpositions or refined quantum correlations such as quantum entanglement. It is, therefore, necessary for a claim of quantum superiority to prove if it is indeed harvesting (erasing) quantum coherence for quantum-enhanced device operations. A simple situation to clarify this point is superradiance. Classical antennas or excited atoms emit radiation scaled quadratically with the number of radiators; hence a simple quadratic scaling with the number of resources in energy processes, storage, or emission, is not a genuine quantum advantage. In particular, most of the models of quantum systems assume qubits or spins coupled with size-independent interaction coefficients. However, in the thermodynamic limit, such interactions may predict superlinear scaling with the number of qubits or spins. Such energy scaling contradicts the principles of extensive thermodynamics, where the energy must scale linearly with the system size. In models of spin-spin or qubit-qubit interactions, when larger system limits are considered, one must take into account the spatial (size) dependence of spin-spin or qubit-qubit interactions carefully simple way to do this is to employ so-called Kac scaling prescription, which divides the coupling coefficients with the number of spins or qubits.

Designing quantum harvesting systems, such as quantum computers or quantum machines (heat engines, batteries, or energy converters, for example) is then crucial to operate first on the proper basis to harness quantum coherence to ensure genuine quantum superlinear advantage. Second, they should maintain a quantum superlinear advantage even if the interactions are subject to Kac scaling or system size dependence. Avoiding reduced coupling with the system size can be more straightforward for dissipatively generated and sustained quantum entanglement and coherence. Shared baths can mediate ubiquitous long-range entanglement or coherence between many qubits even if they are not directly interacting. Long-range coupled systems do not have to be extensive, and hence they do not contradict thermodynamical extensivity principles. M. Scully and co-workers have proposed a proper quantum coherence harvester for work extraction by erasing quantum coherence~\cite{Scully:2003aa}. This pioneering work has been generalized to a higher number of qubits in Ref.~\cite{Turkpence:2016wo} which shows that genuine quantum superiority with quadratic scaling in extracted work and efficiency can be obtained environmental decoherence and dephasing scale slower than the quantum coherence.

Ref.~\cite{Pusuluk:2021wf} considers quantum information reservoirs as non-thermal baths and establishes the general relations between the chemical potential, quantum coherence, temperature differences, and the heat, particle, and information currents. The approach generalizes the usual thermoelectric Onsager relations, and Rayleigh heat conduction Quantum coherent equivalents of thermoelectric Peltier and Seebeck effects have been found. The interplay of quantum discord, coherence, and entanglement with the heat flows is pointed out. Similar to the dielectric classification of materials, the authors introduced the term and concept of ``dicoherent materials'', quantified in terms of their thermocoherent coefficients, the analog of thermoelectric coefficients but measuring the conductance of material for coherence flow. Such materials are critical to designing reservoirs that can protect quantum coherence or entanglement within, especially when the entanglement gets larger, the analog of using dielectrics in capacitors. After a certain number of qubits are entangled, the multiple qubit system becomes a non-equilibrium system relative
to the environment with no quantum coherence or entanglement. Hence, according to the thermocoherent Onsager theory, a coherence flow is expected. Using a dicoherent material between the environment (with no coherence or entanglement) and system (with coherence or entanglement), one can keep more qubits entangled, similar to
avoiding dielectric breakdown in capacitor charging.

Generation of quantum entanglement or coherence in a quantum system using a thermal gradient can be envisioned as a quantum heat engine~\cite{Tavakoli:2018vy,Man:2015we,Tacchino:2018vi}, which is the practical application of emerging field of quantum thermodynamics~\cite{OZDEMIR:2020vj,Kosloff:2013ta}. M. Scully argued that quantum coherence could be exploited to enhance photovoltaic efficiencies~\cite{Scully:2010vo} and efficiency of a photonic Carnot engine~\cite{Scully:2003vt}.

\subsection{Sustained Quantum Entanglement in a Single Engineered Environment}

Can a single, shared environment to multiple qubits be sufficient to generate and sustain quantum entanglement among them? A much simpler version of this question was answered positively first for the case of quantum coherence. G.~Guarnieri and co-workers determined the conditions on how a two-level system must be coupled to a thermal bath to possess quantum coherence in steady state~\cite{Guarnieri:2018wm,Guarnieri:2021wn}. Steady-state coherence is independent of the initial state of the two-level system (qubit). Driven by a single thermal bath, the qubit can still build significant
quantum coherence in a steady state, even if the qubit starts in an incoherent state. The steady-state coherence is maintained for an arbitrary time, as long as the qubit remains coupled to the bath.
The scheme does not need any coherent drives so that sustained quantum superposition is only powered by thermal noise.
Rectification of thermal noise into a more profound and refined quantum state, an entangled multiple qubit state, is a more challenging problem. An early answer to the case of entanglement is presented by M.~B.~Plenio and S.~F.~Huelga, who showed that a purely incoherent, thermal, shared bath is sufficient to entangle photons in two distinct optical cavities~\cite{Plenio:2002wc}. Similar positive answers have been established for two atoms in a shared bath in Ref.~\cite{Benatti:2010us,Guarnieri:2018wm}. A general theoretical study established some conditional relations between quantum entanglement and quantum discord in a three-partite system, where a subsystem mediates interactions between the other two uncoupled subsystems~\cite{Krisnanda:2017uj}. The Quantum character of an elusive, inaccessible, mediator subsystem can be tested with the entanglement of the probe subsystems, which could be applicable to test quantumness of the gravitational field, for example~\cite{Marletto:2017tc,Krisnanda:2017uj}. The conclusions are valid even when the subsystems are coupled to their local environments. Based upon the insight provided by Ref.~\cite{Krisnanda:2017uj}, we can think that the essential physics behind the quantum entanglement induced by a shared environment consists of two main ingredients. First is the quantum character of the shared environment. Second is the quantum discord between the shared entanglement and the partition of the other two subsystems. If a quantum shared environment can build sufficient quantum discord between itself and the partition of the other two quantum objects, then the shared environment can induce quantum entanglement even under local decoherence channels.

A systematic study has revealed which quantum coherences are accessible by thermal, chaotic resources and which are accessible by coherent resources~\cite{Dag:2016vy}. Subsequent analyses were pointed out how an initially independent ensemble of multiple ($N$) qubits can be thermally charged to a quantum coherent state and revealed that the steady-state coherence increases with $N^2$~\cite{Cakmak:2017vz}. However, these conclusions are valid if only shared thermal environments are present. In practice, there are uncontrollable local quantum dephasing or decoherence channels, and in this case, the steady-state coherence or entanglement can be easily destroyed.

Entanglement of quantum systems using thermal noise or thermal entanglement can be considered as a natural route to entanglement  generation~\cite{Arnesen:2001uy}. It is relatively easy to scale thermal entanglement to large systems. Quantum entanglement in an Ising chain using a magnetic field and only a single thermal reservoir has been theoretically shown~\cite{Gunlycke:2001vy}. Even if the other noise sources can eventually destroy quantum coherence, correlations, or entanglement, their presence for sufficiently long enough time in a natural process may make it more efficient, which is believed to be the case by some researchers is the case in light-harvesting complexes. Noise-assisted generation of transient or steady entanglement can, in return, yield entanglement enhanced efficient transport or energy processes. Can nature utilize other means to strengthen entanglement? As we know from earlier research, shared baths, multiple non-equilibrium reservoirs, finite environments can further enhance
the steady-state entanglement. In light-harvesting complexes, the entanglement, coherence, or correlations are only
needed for a sufficiently long time and in a particular localization. The other, more extensive set of atoms from the different parts of the complex molecule can be considered a background matrix, providing a protection layer, making vibrations, and causing positive feedback as a non-Markovian finite bath, to further compensate decoherence
effects from other environments. In addition, the shared thermal bath with those extra contributions of memory or coherence induction properties could sustain the entanglement against local decoherence channels. To verify such a  hypothesis of quantum coherence and correlation generation and protection for a biological system is not easy. Yet the challenge is addressed by some researchers, at least for quantum coherence but not for entanglement, who concluded, based upon their experimental analysis, that the background matrix (protein environment around the bacteriochlorophyll chromophores) in photosynthetic light-harvesting complexes prolongs quantum coherence, which in return enhances energy transport efficiency~\cite{Panitchayangkoon:2011vb}.

Alternatively, one can follow a biomimetic strategy and imitate the steady-state entanglement generation and protection route in a synthetic system. Ref.~\cite{Ullah:2021wa} shows that using a shared magnon bath in a YIG crystal for two distant NV centers in a diamond host located on top of the YIG crystal; one can produce entangled NV centers in steady-state. This entanglement, however, is subject to sudden death if one considers NV centers other environments, nuclear spins in diamond, and electromagnetic radiation modes into which NV centers radiate spontaneously. Ref.~\cite{Ullah:2021wa} shows that such additional decoherence and dephasing channels can be counterbalanced, and entanglement can show sudden birth at
later times by using external electric and magnetic fields on the system. These fields are static and simply allow for engineering the magnon dispersion relation to sculpt the spectral response function of the magnon bath and make the bath non-thermal. The bath mode oscillators are displaced by the magnetic field so that instead of a thermal bath, the magnons become coherent thermal bath, a collection of displaced oscillators. Accordingly, spectral filtering and induced coherence on the magnon bath turn out to be sufficient for
rebirth and sustainability of the quantum entanglement of distant NV centers. The method can be scaled to
multiple NV centers as the distant NV centers do not require to be coupled directly.
The engineered magnon bath acts as a material with a high ``dicoherence constant'' (analog of a capacitor with a high dielectric constant filling medium), isolating the entangled NV centers from other environments without quantum coherence or correlations acting as a sink to destroy entanglement.

\section{Conclusion}

In this short overview, we focused upon the subject of steady-state quantum entanglement and briefly touched on the similar stabilization of quantum coherence. In a broader picture, resetting a quantum system to a target, usually ground, the state is a practical goal for many quantum technology applications, in particular for quantum error correction demanded fault-tolerant quantum computation or for thermal quantum annealing. While the dissipative routes, such as cooling to state generation, are scalable, energy, and hardware efficient, they are naturally slow processes. Recent progress in so-called shortcuts to equilibration~\cite{Dann:2019tb}, extending the success of shortcuts to adiabatic processes, can be promising to make dissipative schemes competitively rapid and more appealing for computing applications. From an energy perspective, dissipative methods to produce and sustain quantum coherence and entanglement open new fields of quantum technology applications, such as quantum photovoltaics and quantum heat or information engines, including thermal diodes transistors and switches for quantum phononic computations. Such dissipative approaches may reduce the energy footprints of quantum technology applications. The possibility of these dissipative schemes of quantum coherence and entanglement generation in biological systems is of fundamental significance and can illuminate our understanding of energy and information processes in biological systems. Without constraints of living organisms, one can also learn from natural processes but design more versatile, efficient, and powerful synthetic devices.

%\subsubsection{Figures}

%Tables are numbered consecutively and should include a clear
%descriptive caption (see Figure \ref{fig-eg}).

%\begin{figure}
%\begin{center}
%\scalebox{0.5}{\includegraphics{union.pdf}}
%\end{center}
%\caption{Example figure, taken from reference \cite{St1}.}
%\label{fig-eg}
%\end{figure}

\section{Acknowledgments}

We gratefully acknowledge Dr. Ned Allen for many fruitful discussions and encouragement to research quantum coherence from energy perspectives. \"{O}.~E.~M.~acknowledges support from TUBITAK Grant
No.~120F230.

\bibliography{ijuc}

%\appendix

\end{document}